\documentstyle[pra,aps]{revtex}

\begin{document}
\title{{\Large Multimode cat-state entanglement and network teleportation}}
\author{{\bf Ba An Nguyen\thanks{%
Corresponding Author. Email: nbaan@kias.re.kr} and Jaewan Kim\thanks{%
Email: jaewan@kias.re.kr}}}
\address{School of Computational Sciences, Korea Institute for Advanced Study, \\
207-43 Cheongryangri-dong, Dongdaemun-gu, Seoul 130-722, Republic of Korea}
\maketitle
\date{}

\begin{abstract}
Schemes for generation and protocols for network teleportation of multimode
entangled cat-states are proposed. Explicit expressions for probability of
successful teleportation are derived for both symmetric and asymmetric
networks.

\end{abstract}


\noindent {\bf 1. Introduction}

\noindent Entanglement as the most characteristic trait of quantum mechanics
is thought to be one of the few main keys to open the door to future
cyber-information development including superdense coding, telecloning,
quantum cryptography, quantum computation, {\it etc.} In particular,
inspired by the more-than-surprising idea by Bennett {\it et al}. \cite{ben}%
, who for the first time proposed a quantum way beautifully based on
entanglement to teleport an unknown qubit between two space-like distant
stations, various other teleportation protocols have been dealt with.
Nowadays pure photonic \cite{pho}, atomic \cite{ato} states as well as
entangled states \cite{ent} and states with continuous variables \cite{con}
can be teleported successfully. Schr\"{o}dinger cat-states \cite{sch} can
also be teleported employing entangled coherent states \cite{ecs} as a
quantum channel. Recently, it turns out that single qubits and entangled
qubits can be encoded using cat-states of bosonic fields \cite{chuang}. The
bosonic modes could be of optical nature \cite{opt} or of vibrational motion
of trapped ions \cite{ion}, {\it etc}. As an example, logical states for
quantum computation could be encoded in terms of even and odd coherent
states with the advantage that these cat-states are sharply distinguishable
by their number parity. So, bit-flip errors, if any, are easily to be
detected and corrected by means of a proper circuit \cite{corr}. As for
quantum information, protocols for teleporting cat-states have been proposed
as well using pure \cite{enk,wang,zheng} or mixed \cite{jeong} quantum
channels which are again types of (entangled) cat-states. The cat-state to
be teleported is single-mode and two-mode in \cite{enk,jeong,zheng} and \cite
{wang}, respectively. Nevertheless, only two parties, Alice and Bob, have
been engaged in the game in \cite{enk,wang,zheng,jeong}. Concerning
practical purposes, information communication should be realizable within a
network consisting of as many parties as possible. Such a network problem
has been addressed in connection with cloning machines \cite{clon} as well
as with teleportation scenarios \cite{tele1,tele2}. Note that in \cite{tele1}
three players were involved in the game with a discrete two-state system,
while in \cite{tele2} a general network was considered for teleporting
quadratures of a continuous field. The aim of our work here is to construct
suitable quantum channels to teleport an unknown multimode cat-state within
a network with any number $N$ of participants. The cat-state to be
teleported is of the general multimode form 
\begin{equation}
\left| \Psi \right\rangle _{a_{1}a_{2}...a_{L}}=A_{a_{1}a_{2}...a_{L}}\left(
x\left| \alpha \right\rangle _{a_{1}}\left| \alpha \right\rangle
_{a_{2}}...\left| \alpha \right\rangle _{a_{L}}+y\left| -\alpha
\right\rangle _{a_{1}}\left| -\alpha \right\rangle _{a_{2}}...\left| -\alpha
\right\rangle _{a_{L}}\right) ,  \label{E1}
\end{equation}
where 
\begin{equation}
\left| \alpha \right\rangle _{a_{j}}=\exp [-|\alpha
|^{2}/2]\sum_{n=0}^{\infty }\frac{\alpha ^{n}}{n!}(a_{j}^{+})^{n}\left|
0\right\rangle _{j},
\end{equation}
with $\alpha \in {\cal C}$, $a_{j}^{+}$ $(a_{j})$ and $\left| 0\right\rangle
_{j}$ the photon creation (annihilation) operator and the vacuum of mode $j,$
is a coherent state, 
\begin{equation}
A_{a_{1}a_{2}...a_{L}}=\left( |x|^{2}+|y|^{2}+2\exp (-2L|\alpha |^{2})\Re
(x^{*}y)\right) ^{-1/2}
\end{equation}
is the normalization coefficient, $L=1,2,\ldots $ and $x,y$ are unknown
complex amplitudes. We consider two types of networks: symmetric and
asymmetric. By symmetric network we mean a network all parties of which are
equivalent to each other. That is, any one can hold the state to be
teleported and any another one is able to receive it. The quantum channel
can be shared equally among the parties before an actual action of
teleportation takes place and therefore teleportation can be performed
between any pair of parties within the symmetric network. By asymmetric
network we mean a network in which one party who is in possession with the
state to be teleported is distinguished from the remaining parties. The
quantum channel cannot be shared by the parties until a decision is made of
who is the teleporting party, i.e. who holds the state to be teleported.
After this is decided the parties start to share the quantum channel but the
sharing is unequal among the participants. Teleportation can go only from
one certain party to another party in the asymmetric network. The structure
of the cat-state to be teleported, state (\ref{E1}), suggests that we use as
a quantum channel multimode entangled cat-states (also called multimode
even/odd entangled coherent states) of the form \cite{ansari} 
\begin{equation}
\left| \Psi _{M}^{\pm }\right\rangle =A_{M}^{\pm }\left| \widetilde{\Psi }%
_{M}^{\pm }\right\rangle ,  \label{e12M}
\end{equation}
with $M=2,3,\ldots ,$

\begin{equation}
\left| \widetilde{\Psi }_{M}^{\pm }\right\rangle =\left| \Phi
_{M}^{+}\right\rangle \pm \left| \Phi _{M}^{-}\right\rangle ,
\end{equation}
\begin{equation}
\left| \Phi _{M}^{\pm }\right\rangle =\left| \pm \alpha \right\rangle
_{1}\left| \pm \alpha \right\rangle _{2}...\left| \pm \alpha \right\rangle
_{M},
\end{equation}
and 
\begin{equation}
A_{M}^{\pm }=\left[ 2\left( 1\pm \exp (-2M|\alpha |^{2})\right) \right]
^{-1/2}
\end{equation}
the normalization coefficients.

We divide our paper into Introduction and three more sections. In section 2
we prepare the general multimode entangled cat-states by two schemes using
only linear optical devices combined with simple parity measurements. These
states are to be served as a quantum channel in section 3 to teleport an
unknown state $\left| \Psi \right\rangle _{a_{1}a_{2}...a_{L}}.$ Relevant
discussions are provided in each section and the paper ends up with a
conclusion, the final section.

\noindent {\bf 2. Generation schemes}

\noindent In this section we propose two schemes to generate the multimode
entangled cat-states $\left| \Psi _{M}^{\pm }\right\rangle $ with an
arbitrary integer $M\geq 2,$ given a single-mode odd/even coherent state
(which is easily produced from a single-mode coherent state by means of a
Kerr-medium \cite{eocs}). In the first scheme only linear optics is needed
while in the second scheme some measurements should be accompanied.

As is well-known, a lossless beam-splitter is described by the unitary
operator 
\begin{equation}
\widehat{b}_{k,l}(\theta )=\exp [i\theta (a_{k}^{+}a_{l}+a_{l}^{+}a_{k})]
\end{equation}
and a phase-shifter by 
\begin{equation}
\widehat{P}_{j}(\varphi )=\exp (-i\varphi a_{j}^{+}a_{j}).
\end{equation}
We shall use two types of modified beam-splitters defined as follows

\begin{equation}
\widehat{B}_{k,l}(\theta )\equiv \widehat{b}_{k,l}(\theta )\widehat{P}%
_{l}(\pi /2),  \label{B}
\end{equation}

\begin{equation}
\widehat{{\cal B}}_{k,l}\equiv \widehat{P}_{l}(\pi /2)\widehat{b}_{k,l}(\pi
/4)\widehat{P}_{l}(\pi /2).  \label{BN}
\end{equation}
They act on coherent states as

\begin{equation}
\widehat{B}_{k,l}(\theta )\left| \alpha \right\rangle _{k}\left| \beta
\right\rangle _{l}=\left| \alpha \cos \theta +i\beta \sin \theta
\right\rangle _{k}\left| \alpha \sin \theta -i\beta \cos \theta
\right\rangle _{l},  \label{BB}
\end{equation}

\begin{equation}
\widehat{{\cal B}}_{k,l}\left| \alpha \right\rangle _{k}\left| \beta
\right\rangle _{l}=\left| (\alpha +\beta )/\sqrt{2}\right\rangle _{k}\left|
(\alpha -\beta )/\sqrt{2}\right\rangle _{l}.  \label{BH}
\end{equation}
While the first generation scheme uses operators $\widehat{B}_{k,l}(\theta
), $ the second one is based on operators $\widehat{{\cal B}}_{k,l}.$ By
reasons that will be made clear later, the first scheme is referred to as
Ladder-scheme and the second one as Tree-scheme.

\noindent {\it 2.1. Ladder-scheme}

\noindent It can be proved by induction that

\begin{equation}
\prod_{q=M-1}^{1}\widehat{B}_{q,q+1}(\theta _{M+1-q})\left| \pm \alpha \sqrt{%
M}\right\rangle _{1}\left| 0\right\rangle _{2}...\left| 0\right\rangle
_{M}=\left| \Phi _{M}^{\pm }\right\rangle  \label{a1}
\end{equation}
where 
\begin{equation}
\theta _{n}=\cos ^{-1}\frac{1}{\sqrt{n}}.  \label{theta}
\end{equation}

Indeed, for $M=2,$ Eq. (\ref{a1}) reduces at once to Eq. (\ref{BB}) with the
replacements $\alpha \rightarrow \pm \alpha \sqrt{2},$ $\beta \rightarrow 0.$
We assume Eq. (\ref{a1}) being true for a given $M>2$ and try to prove that

\begin{equation}
\prod_{q=M}^{1}\widehat{B}_{q,q+1}(\theta _{M+2-q})\left| \pm \alpha \sqrt{%
M+1}\right\rangle _{1}\left| 0\right\rangle _{2}...\left| 0\right\rangle
_{M+1}=\left| \Phi _{M+1}^{\pm }\right\rangle  \label{a2}
\end{equation}
is also true. The l.h.s. of Eq. (\ref{a2}) can be rewritten as

\begin{equation}
\prod_{q=M}^{2}\widehat{B}_{q,q+1}(\theta _{M+2-q})\widehat{B}_{1,2}(\theta
_{M+1})\left| \pm \alpha \sqrt{M+1}\right\rangle _{1}\left| 0\right\rangle
_{2}...\left| 0\right\rangle _{M+1}
\end{equation}
which by virtue of Eqs. (\ref{BB}) and (\ref{theta}) becomes

\begin{equation}
\left| \pm \alpha \right\rangle _{1}\prod_{q=M}^{2}\widehat{B}%
_{q,q+1}(\theta _{M+2-q})\left| \pm \alpha \sqrt{M}\right\rangle _{2}\left|
0\right\rangle _{3}...\left| 0\right\rangle _{M+1}.  \label{a4}
\end{equation}
If in Eq. (\ref{a1}) a change of mode labelling $q\rightarrow q+1$ is made
this equation will look like

\begin{equation}
\prod_{q=M}^{2}\widehat{B}_{q,q+1}(\theta _{M+2-q})\left| \pm \alpha \sqrt{M}%
\right\rangle _{2}\left| 0\right\rangle _{3}...\left| 0\right\rangle
_{M+1}=\left| \pm \alpha \right\rangle _{2}\left| \pm \alpha \right\rangle
_{3}...\left| \pm \alpha \right\rangle _{M+1}.  \label{a3}
\end{equation}
Taking Eq. (\ref{a3}) into account in Eq. (\ref{a4}) proves the correctness
of Eq. (\ref{a2}). The desired states $\left| \Psi _{M}^{\pm }\right\rangle $
are thus generated as

\begin{equation}
\left| \Psi _{M}^{\pm }\right\rangle =A_{M}^{\pm }\widehat{B}_{M}\left(
\left| \alpha \sqrt{M}\right\rangle _{1}\pm \left| -\alpha \sqrt{M}%
\right\rangle _{1}\right) \left| 0\right\rangle _{2}\left| 0\right\rangle
_{3}...\left| 0\right\rangle _{M}  \label{Bsc}
\end{equation}
where the operator $\widehat{B}_{M}$ is defined by

\begin{equation}
\widehat{B}_{M}\equiv \prod_{q=M-1}^{1}\widehat{B}_{q,q+1}(\theta _{M+1-q}).
\label{BM}
\end{equation}
The schema of arrangement of beam-splitters $\widehat{B}_{q,q+1}(\theta
_{M+1-q}),$ Fig. 1 a, looks like a ladder leading to the name Ladder-scheme.

\noindent {\it 2.2. Tree-scheme}

\noindent Suppose that we have the state

\begin{equation}
\left| \pm \alpha \sqrt{2}\right\rangle _{1}\left| \pm \alpha \sqrt{2}%
\right\rangle _{2}...\left| \pm \alpha \sqrt{2}\right\rangle
_{2^{Q-1}}\left| 0\right\rangle _{1+2^{Q-1}}\left| 0\right\rangle
_{2+2^{Q-1}}...\left| 0\right\rangle _{2^{Q}}\equiv
\prod_{j=1}^{2^{Q-1}}\left| \pm \alpha \sqrt{2}\right\rangle
_{j}\prod_{k=1}^{2^{Q-1}}\left| 0\right\rangle _{k+2^{Q-1}}  \label{b1}
\end{equation}
with $Q$ a positive integer. An action of the sequence of operators

\begin{equation}
\widehat{{\cal B}}_{1,1+2^{Q-1}}\widehat{{\cal B}}_{2,2+2^{Q-1}}...\widehat{%
{\cal B}}_{2^{Q-1},2^{Q}}\equiv \prod_{q=1}^{2^{Q-1}}\widehat{{\cal B}}%
_{q,q+2^{Q-1}}
\end{equation}
on the state (\ref{b1}), on account of Eq. (\ref{BH}), yields the states $%
\left| \Phi _{2^{Q}}^{\pm }\right\rangle ,$ {\it i.e.}

\begin{equation}
\prod_{q=1}^{2^{Q-1}}\widehat{{\cal B}}_{q,q+2^{Q-1}}\prod_{j=1}^{2^{Q-1}}%
\left| \pm \alpha \sqrt{2}\right\rangle _{j}\prod_{k=1}^{2^{Q-1}}\left|
0\right\rangle _{k+2^{Q-1}}=\left| \Phi _{2^{Q}}^{\pm }\right\rangle .
\label{b2}
\end{equation}
Repeated use of the above equation in its l.h.s. eventuates in

\begin{equation}
\widehat{{\cal B}}_{2^{Q}}\left| \pm \alpha \sqrt{2^{Q}}\right\rangle
_{1}\prod_{k=2}^{2^{Q}}\left| 0\right\rangle _{k}=\left| \Phi _{2^{Q}}^{\pm
}\right\rangle  \label{b3}
\end{equation}
where the operator $\widehat{{\cal B}}_{2^{Q}}$ is defined by

\begin{equation}
\widehat{{\cal B}}_{2^{Q}}\equiv \prod_{l=Q}^{1}\left( \prod_{q=1}^{2^{l-1}}%
\widehat{{\cal B}}_{q,q+2^{l-1}}\right) .  \label{B2Q}
\end{equation}
The state $\left| \Psi _{M}^{\pm }\right\rangle $ with $M=2^{Q}$ can thus be
generated as

\begin{equation}
\left| \Psi _{M=2^{Q}}^{\pm }\right\rangle =A_{2^{Q}}^{\pm }\widehat{{\cal B}%
}_{2^{Q}}\left( \left| \alpha \sqrt{2^{Q}}\right\rangle _{1}\pm \left|
-\alpha \sqrt{2^{Q}}\right\rangle _{1}\right) \prod_{k=2}^{2^{Q}}\left|
0\right\rangle _{k}.  \label{S2Q}
\end{equation}
The schema of arrangement of beam-splitters $\widehat{{\cal B}}%
_{q,q+2^{l-1}},$ Fig. 1 b, resembles a tree suggesting the name Tree-scheme.

For $M\neq 2^{Q}$ additional measurements are needed. To see what kinds of
measurement are required let us suppose that we have the states $\left| 
\widetilde{\Psi }_{K}^{\pm }\right\rangle $ with $K=2,3,...$ It is a
straightforward matter to verify the following equality

\begin{equation}
\left| \widetilde{\Psi }_{K}^{\pm }\right\rangle =\frac{1}{2}\left( \left| 
\widetilde{\Psi }_{K-1}^{-}\right\rangle \left| \mp \right\rangle
_{K}+\left| \widetilde{\Psi }_{K-1}^{+}\right\rangle \left| \pm
\right\rangle _{K}\right)  \label{b4}
\end{equation}
where

\begin{equation}
\left| \pm \right\rangle _{K}=\left| \alpha \right\rangle _{K}\pm \left|
-\alpha \right\rangle _{K}.
\end{equation}
Repeatedly applying Eq. (\ref{b4}) to its r.h.s. we have established the
following general formula valid for an integer $J\in [1,K]$

\begin{equation}
\left| \widetilde{\Psi }_{K}^{\pm }\right\rangle =\left| \widetilde{\Psi }%
_{K-J}^{-}\right\rangle \left| F_{J}^{\mp }\right\rangle +\left| \widetilde{%
\Psi }_{K-J}^{+}\right\rangle \left| F_{J}^{\pm }\right\rangle .  \label{b5}
\end{equation}
In Eq. (\ref{b5}) the $J$-mode states $\left| F_{J}^{\pm }\right\rangle $
are determined recurrently as

\begin{equation}
\left| F_{J}^{\pm }\right\rangle =\frac{1}{2}\left( \left| +\right\rangle
_{K-J+1}\left| F_{J-1}^{\pm }\right\rangle +\left| -\right\rangle
_{K-J+1}\left| F_{J-1}^{\mp }\right\rangle \right)  \label{b6}
\end{equation}
with the conventions $\left| F_{0}^{+}\right\rangle =1$ and $\left|
F_{0}^{-}\right\rangle =0.$

The state $\left| +\right\rangle _{j}$ $(\left| -\right\rangle _{j})$ has
parity $\pi _{j}=+1$ $(-1)$ because it contains only an even (odd) number of
photons. It is not difficult to realize from Eq. (\ref{b6}) that all the
states $\left| F_{k}^{+}\right\rangle $ $(\left| F_{k}^{-}\right\rangle )$
with $k>0$ have a positive (negative) parity. The cruciality of Eq. (\ref{b5}%
) rests in the fact that having $\left| \widetilde{\Psi }_{K}^{\pm
}\right\rangle $ we are able to generate $\left| \widetilde{\Psi }%
_{K-J}^{+}\right\rangle $ or $\left| \widetilde{\Psi }_{K-J}^{-}\right%
\rangle $ by state-parity measurements to be described below. Let us have $%
\left| \widetilde{\Psi }_{K}^{-}\right\rangle $ and we wish to generate $%
\left| \widetilde{\Psi }_{K-J}^{-}\right\rangle $ $(\left| \widetilde{\Psi }%
_{K-J}^{+}\right\rangle ).$ For that purpose, we measure state-parities of $%
J $ modes $\{K-J+1,$ $K-J+2,$ ..., $K\}.$ If the outcome gives a total
parity $\prod_{k=1}^{J}\pi _{K-J+k}=+1$ $(-1),$ then the collapsed state is
the desired $\left| \widetilde{\Psi }_{K-J}^{-}\right\rangle $ $(\left| 
\widetilde{\Psi }_{K-J}^{+}\right\rangle ).$ On the contrary, if we start
from $\left| \widetilde{\Psi }_{K}^{+}\right\rangle ,$ then the outcome $%
\prod_{k=1}^{J}\pi _{K-J+k}=+1$ $(-1)$ means generation of the state $\left| 
\widetilde{\Psi }_{K-J}^{+}\right\rangle $ $(\left| \widetilde{\Psi }%
_{K-J}^{-}\right\rangle )$ instead. We now exploit Eq. (\ref{b5}) to
generate states $\left| \Psi _{M}^{\pm }\right\rangle $ when $M$ is not a
power of $2.$ In this case, there always exist integers $Q_{j}$ such that $%
M<2^{Q_{j}}.$ Among $Q_{j}$ let $Q$ be the smallest one, i.e. $Q=\min
\{Q_{j}\}.$ To generate $\left| \Psi _{M}^{-}\right\rangle $ we first act $%
\widehat{{\cal B}}_{2^{Q}}$ (see the definition (\ref{B2Q})) on $%
A_{2^{Q}}^{+}\left( \left| \alpha \sqrt{2^{Q}}\right\rangle _{1}+\left|
-\alpha \sqrt{2^{Q}}\right\rangle _{1}\right) \prod_{k=2}^{2^{Q}}\left|
0\right\rangle _{k}$ to prepare the state $\left| \Psi
_{2^{Q}}^{+}\right\rangle =A_{2^{Q}}^{+}\left| \widetilde{\Psi }%
_{2^{Q}}^{+}\right\rangle $ which, on account of Eq. (\ref{b5}), reads

\begin{equation}
A_{2^{Q}}^{+}\left( \left| \widetilde{\Psi }_{M}^{-}\right\rangle \left|
F_{2^{Q}-M}^{-}\right\rangle +\left| \widetilde{\Psi }_{M}^{+}\right\rangle
\left| F_{2^{Q}-M}^{+}\right\rangle \right) .  \label{b7}
\end{equation}
Then we measure the total state-parity of modes $M+1,$ $M+2,$ $...,$ $2^{Q}.$
If the total parity $\prod_{k=1}^{2^{Q}-M}\pi _{M+k}=-1$ the state (\ref{b7}%
) collapses into the desired state $\left| \Psi _{M}^{-}\right\rangle .$ The
probability of successful generation is derived as

\begin{eqnarray}
P_{Q,M}^{+-} &=&\left( \frac{A_{2^{Q}}^{+}}{A_{M}^{-}}\right) ^{2}\left|
_{2^{Q}}\left\langle n_{2^{Q}}\right| ..._{M+2}\left\langle n_{M+2}\right|
_{M+1}\left\langle n_{M+1}\right| \left. F_{2^{Q}-M}^{-}\right\rangle
\right| ^{2}  \nonumber \\
&=&\left( \frac{A_{2^{Q}}^{+}}{A_{M}^{-}}\right) ^{2}\sum_{\{n_{j}=0\},\sum
n_{j}=odd}^{\infty }\left| {\cal N}_{n_{M+1}}(\alpha ){\cal N}%
_{n_{M+2}}(\alpha )...{\cal N}_{n_{2^{Q}}}(\alpha )\right| ^{2}  \nonumber \\
&=&\frac{(1-\text{e}^{-2M|\alpha |^{2}})}{2(1+\text{e}^{-2^{(Q+1)}|\alpha
|^{2}})(1-\text{e}^{-2(2^{Q}-M)|\alpha |^{2}})}.  \label{Pct}
\end{eqnarray}
In Eq. (\ref{Pct}) 
\begin{equation}
{\cal N}_{n}(\beta )=\text{ }_{k}\left\langle n\right| \left. \beta
\right\rangle _{k}=\exp (-|\beta |^{2}/2)\beta ^{n}/\sqrt{n!}  \label{NH}
\end{equation}
is mode-independent. Alternatively, if we act $\widehat{{\cal B}}_{2^{Q}}$
on $A_{2^{Q}}^{-}\left( \left| \alpha \sqrt{2^{Q}}\right\rangle _{1}-\left|
-\alpha \sqrt{2^{Q}}\right\rangle _{1}\right) \prod_{k=2}^{2^{Q}}\left|
0\right\rangle _{k}$ to prepare the state $\left| \Psi
_{2^{Q}}^{-}\right\rangle =A_{2^{Q}}^{-}\left| \widetilde{\Psi }%
_{2^{Q}}^{-}\right\rangle $ and perform the same state-parity measurements,
then the outcome $\prod_{k=1}^{2^{Q}-M}\pi _{M+k}=+1$ (not $-1)$ means
generation of the state $\left| \Psi _{M}^{-}\right\rangle $ with the
probability of success equal to

\begin{eqnarray}
P_{Q,M}^{--} &=&\left( \frac{A_{2^{Q}}^{-}}{A_{M}^{-}}\right) ^{2}\left|
_{2^{Q}}\left\langle n_{2^{Q}}\right| ..._{M+2}\left\langle n_{M+2}\right|
_{M+1}\left\langle n_{M+1}\right| \left. F_{2^{Q}-M}^{+}\right\rangle
\right| ^{2}  \nonumber \\
&=&\left( \frac{A_{2^{Q}}^{-}}{A_{M}^{-}}\right) ^{2}\sum_{\{n_{j}=0\},\sum
n_{j}=even}^{\infty }\left| {\cal N}_{n_{M+1}}(\alpha ){\cal N}%
_{n_{M+2}}(\alpha )...{\cal N}_{n_{2^{Q}}}(\alpha )\right| ^{2}  \nonumber \\
&=&\frac{(1-\text{e}^{-2M|\alpha |^{2}})}{2(1-\text{e}^{-2^{(Q+1)}|\alpha
|^{2}})(1+\text{e}^{-2(2^{Q}-M)|\alpha |^{2}})}.  \label{Ptt}
\end{eqnarray}
Similarly, the state $\left| \Psi _{M}^{+}\right\rangle $ can be generated
either from $\left| \Psi _{2^{Q}}^{+}\right\rangle $ or from $\left| \Psi
_{2^{Q}}^{-}\right\rangle $ with the success probabilities given
respectively by

\begin{equation}
P_{Q,M}^{++}=\coth (M|\alpha |^{2})\tanh (2^{Q}|\alpha |^{2})P_{Q,M}^{--}
\label{Pcc}
\end{equation}
and

\begin{equation}
P_{Q,M}^{-+}=\coth (M|\alpha |^{2})\coth (2^{Q}|\alpha |^{2})P_{Q,M}^{+-}.
\label{Ptc}
\end{equation}
When $|\alpha |\rightarrow \infty $ all the probabilities are approaching $%
0.5$ but in the limit $|\alpha |\rightarrow 0$ their behaviors are
different, 
\begin{equation}
\lim_{|\alpha |\rightarrow 0}P_{Q,M}^{+-}=0,\text{ }\lim_{|\alpha
|\rightarrow 0}P_{Q,M}^{--}=M/2^{Q},\text{ }\lim_{|\alpha |\rightarrow
0}P_{Q,M}^{++}=1,\text{ }\lim_{|\alpha |\rightarrow 0}P_{Q,M}^{-+}=1-M/2^{Q},
\label{lim}
\end{equation}
as seen from Fig. 2 where the probabilities are plotted as a function of $%
|\alpha |^{2}$ for a fixed $Q$ and several values of $M.$ The figure clearly
indicates a preference of $\left| \Psi _{2^{Q}}^{\pm }\right\rangle $ to $%
\left| \Psi _{2^{Q}}^{\mp }\right\rangle $ in the process of generating $%
\left| \Psi _{M}^{\pm }\right\rangle .$ Moreover, the probabilities $%
P_{Q,M}^{--}$ and $P_{Q,M}^{++}$ are always greater than or equal to 50\%
and both of them are larger for smaller difference between $2^{Q}$ and $M.$

\noindent {\it 2.3. Discussion}

\noindent The operator $\widehat{B}_{M}$ defined in Eq. (\ref{BM}) has also
been applied \cite{tele2} to produce generalized multiparticle EPR states 
\cite{epr}, {\it i.e.} eigenstates with zero total momentum and zero
relative positions. The ladder-scheme looks quite simple through Eq. (\ref
{Bsc}). Construction of the operator $\widehat{B}_{M}$ defined by Eq. (\ref
{BM}) is however difficult practically since all the involved beam-splitters
differ one from another: each has to have its own carefully prepared
parameter $\theta _{n}$ depending delicately on the transmissivity and
reflectivity. The tree-scheme, on the other hand, requires only one type of
beam-splitters, the 50:50 ones, which are commonly available. In case $%
M=2^{Q}$ the generation is as simple as via Eq. (\ref{S2Q}). Otherwise,
additional measurements are necessary but these are of the state-parity kind
which can easily performed, say, by employing QED with dispersive atom-field
interactions (see, {\it e.g.}, Ref. \cite{scully}). Although the probability
of success is not 100\%, it is always above 50\% no matter how ``far'' $M$
is from the smallest $2^{Q},$ if starting from the right state ({\it i.e.} $%
\left| \Psi _{2^{Q}}^{\pm }\right\rangle \rightarrow \left| \Psi _{M}^{\pm
}\right\rangle ,$ but not $\left| \Psi _{2^{Q}}^{\pm }\right\rangle
\rightarrow \left| \Psi _{M}^{\mp }\right\rangle ).$ Recently, similar but
related to the center-of-mass vibrational modes of trapped ions multipartite
entangled coherent states have been considered \cite{wangsanders} (see also 
\cite{parkins}). The generation scheme in \cite{wangsanders} is based on
entanglement swapping but the proposed generalized Bell-state measurement of
electronic states seems so challenging even by means of controlled-NOT
gates. Also, the motional entangled cat-states are not relevant for the
teleportation task considered here. As for optical fields, multimode
entangled cat-states may be obtained via the Raman atom-field interaction 
\cite{gerry}, but the entangled modes are confined in many separate cavities
and not ready for distribution among distant users in free space.

Furthermore, it turns out that $\left| \Psi _{M}^{\pm }\right\rangle $ obey
the eigen-equation

\begin{equation}
a_{k_{1}}^{2}a_{k_{2}}^{2}...a_{k_{K}}^{2}\left| \Psi _{M}^{\pm
}\right\rangle =\alpha ^{2K}\left| \Psi _{M}^{\pm }\right\rangle  \label{aa}
\end{equation}
for $1\leq K\leq M.$ In particular, $K=M=3$ corresponds to a special case of
the so-called even/odd trio coherent states \cite{an} which have been shown
genuine nonclassical states. The states $\left| \Psi _{M}^{\pm
}\right\rangle $ therefore intrinsically possess various types of
nonclassical properties that may be exploited for practical purposes (see, 
{\it e.g}., Ref. \cite{weakforce} for weak-force detection). The total
averaged number of photons in the states $\left| \Psi _{M}^{\pm
}\right\rangle $ equals $M\overline{n}^{\pm }$ with

\begin{equation}
\overline{n}^{\pm }=|\alpha |^{2}\times \left\{ 
\begin{tabular}{l}
$\tanh (M|\alpha |^{2})$ \\ 
$\coth (M|\alpha |^{2})$%
\end{tabular}
\right.  \label{ntbct}
\end{equation}
the averaged number in a mode which is mode-independent. At small values of $%
|\alpha |$ there are departures of $\overline{n}^{\pm }$ from $|\alpha
|^{2}, $ the averaged number in a usual single-mode coherent state, but
already for $|\alpha |^{2}\geq 2$ the averaged number per mode, {\it i.e.} $%
\overline{n}^{\pm },$ become coincident with $|\alpha |^{2}$ (see Fig. 3).
An actual total photon number that can be found in $\left| \Psi
_{M}^{+}\right\rangle $ and $\left| \Psi _{M}^{-}\right\rangle ,$ for any $%
M, $ is respectively even and odd, an intuitive physical reason explaining
why $\left\langle \Psi _{M}^{+}\right. \left| \Psi _{M}^{-}\right\rangle =0$
(mathematically such an orthogonality comes directly from the definition of $%
\left| \Psi _{M}^{\pm }\right\rangle $ in Eq. (\ref{e12M})).

A curiosity, as was noticed in \cite{wangsanders}, concerns the two limiting
situations: $|\alpha |\rightarrow \infty $ and $|\alpha |\rightarrow 0.$
When $|\alpha |\rightarrow \infty $ a qubit can be encoded as $\left| \alpha
\right\rangle _{j}=\left| 1_{L}\right\rangle _{j}$ and $\left| -\alpha
\right\rangle _{j}=\left| 0_{L}\right\rangle _{j}$ where $``L"$ stands for
``Logical'' (Note that in this limit $\left| \alpha \right\rangle _{j}$ and $%
\left| -\alpha \right\rangle _{j}$ become orthogonal as $\left\langle \alpha
\right. \left| -\alpha \right\rangle _{j}=\exp (-2|\alpha |^{2})$ vanishes).
Then

\begin{equation}
\left| \Psi _{M}^{\pm }\right\rangle \rightarrow \frac{1}{\sqrt{2}}\left(
\left| 1_{L}\right\rangle _{1}\left| 1_{L}\right\rangle _{2}...\left|
1_{L}\right\rangle _{M}\pm \left| 0_{L}\right\rangle _{1}\left|
0_{L}\right\rangle _{2}...\left| 0_{L}\right\rangle _{M}\right)  \label{GHZ}
\end{equation}
represent the ``{\it all or none}'' states, {\it i.e.} the GHZ-states \cite
{ghz}. Oppositely, $|\alpha |\rightarrow 0$ causes $\left| \Psi
_{M}^{+}\right\rangle \rightarrow 0$ but

\begin{equation}
\left| \Psi _{M}^{-}\right\rangle \rightarrow \frac{1}{\sqrt{M}}\left(
\left| 1_{F}\right\rangle _{1}\left| 0_{F}\right\rangle _{2}...\left|
0_{F}\right\rangle _{M}+\left| 0_{F}\right\rangle _{1}\left|
1_{F}\right\rangle _{2}...\left| 0_{F}\right\rangle _{M}+\left|
0_{F}\right\rangle _{1}\left| 0_{F}\right\rangle _{2}...\left|
1_{F}\right\rangle _{M}\right)  \label{w}
\end{equation}
where $``F"$ stands for ``Fock state''. This is strictly due to entanglement
between modes since in a factorizable case an individual mode $\left| \alpha
\right\rangle _{j}$ tends to $\left| 0_{F}\right\rangle _{j}$ resulting in a
multimode vacuum only. The state (\ref{w}) corresponds to the ``{\it one for
each}'' state, {\it i.e.} the W-state \cite{w}. The above features seem to
bridge between GHZ- and W-state. Yet, these do not at all mean that GHZ- and
W-state are equivalent because they are being approached in two extreme
limits, that cannot be realized by means of local operations and classical
communication, in agreement with what discovered in Ref. \cite{w}.

The bipartite entanglement degree of the state $\left| \Psi _{M}^{\pm
}\right\rangle $ can be assessed by an entanglement measure called
concurrence \cite{concu}. The concurrence between two subsystems: subsystem $%
\{K\}$ consisting of modes $\{j_{1},$ $j_{2},$ ..., $j_{K}\}$ with $%
j_{k}\neq j_{l}\in [1,M],$ $1\leq K<M$, and subsystem $\{M-K\}$ consisting
of the remaining modes $\{j_{K+1},$ $j_{K+2},$ $\ldots $, $j_{M}\},$ can be
calculated as follows. Each subsystem is assumed being linearly independent
with respect to $\alpha $ and $-\alpha $ spanning a two-dimensional
sub-space of the Hilbert space. According to the Gram-Schmidt theorem, we
can always build in each sub-space an orthonormal basis $\{\left|
0\right\rangle _{i},\left| 1\right\rangle _{i}\}$, $i=\{K\},\{M-K\},$ which
is determined as

\begin{equation}
\left| 0\right\rangle _{\{K\}}=\left| \alpha \right\rangle _{j_{1}}...\left|
\alpha \right\rangle _{j_{K}},  \label{x0}
\end{equation}
\begin{equation}
\left| 1\right\rangle _{\{K\}}=\frac{\left| -\alpha \right\rangle
_{j_{1}}...\left| -\alpha \right\rangle _{j_{K}}-\text{ }_{j_{K}}\left%
\langle \alpha \right| ..._{j_{1}}\left\langle \alpha \right| \left. -\alpha
\right\rangle _{j_{1}}...\left| -\alpha \right\rangle _{j_{K}}\left|
0\right\rangle _{\{K\}}}{\sqrt{1-(_{j_{K}}\left\langle \alpha \right|
..._{j_{1}}\left\langle \alpha \right| \left. -\alpha \right\rangle
_{j_{1}}...\left| -\alpha \right\rangle _{j_{K}})^{2}}},  \label{x}
\end{equation}
\begin{equation}
\left| 0\right\rangle _{\{M-K\}}=\left| \alpha \right\rangle
_{j_{K+1}}...\left| \alpha \right\rangle _{j_{M}},
\end{equation}
\begin{equation}
\left| 1\right\rangle _{\{M-K\}}=\frac{\left| -\alpha \right\rangle
_{j_{K+1}}...\left| -\alpha \right\rangle _{j_{M}}-\text{ }%
_{j_{M}}\left\langle \alpha \right| ..._{j_{K+1}}\left\langle \alpha \right|
\left. -\alpha \right\rangle _{j_{K+1}}...\left| -\alpha \right\rangle
_{j_{M}}\left| 0\right\rangle _{\{M-K\}}}{\sqrt{1-(_{j_{M}}\left\langle
\alpha \right| ..._{j_{K+1}}\left\langle \alpha \right| \left. -\alpha
\right\rangle _{j_{K+1}}...\left| -\alpha \right\rangle _{j_{M}})^{2}}}.
\label{x2}
\end{equation}
In terms of the orthonormal bases $\{\left| 0\right\rangle
_{\{K\},\{M-K\}},\left| 1\right\rangle _{\{K\},\{M-K\}}\}$ defined by (\ref
{x0}) - (\ref{x2}), our states $\left| \Psi _{M}^{\pm }\right\rangle $ are
expressed as 
\begin{eqnarray}
\left| \Psi _{M}^{\pm }\right\rangle &=&a_{00}^{\pm }\left| 0\right\rangle
_{\{K\}}\left| 0\right\rangle _{\{M-K\}}+a_{01}^{\pm }\left| 0\right\rangle
_{\{K\}}\left| 1\right\rangle _{\{M-K\}}  \nonumber \\
&&+a_{10}^{\pm }\left| 1\right\rangle _{\{K\}}\left| 0\right\rangle
_{\{M-K\}}+a_{11}^{\pm }\left| 1\right\rangle _{\{K\}}\left| 1\right\rangle
_{\{M-K\}}  \label{x4}
\end{eqnarray}
where 
\begin{eqnarray}
a_{00}^{\pm } &=&A_{M}^{\pm }\left( 1\pm Z^{M}\right) ,  \label{a00} \\
a_{01}^{\pm } &=&\pm A_{M}^{\pm }Z^{K}\sqrt{\left( 1-Z^{2(M-K)}\right) },
\label{a01} \\
a_{10}^{\pm } &=&\pm A_{M}^{\pm }Z^{M-K}\sqrt{\left( 1-Z^{2K}\right) },
\label{a10} \\
a_{11}^{\pm } &=&\pm A_{M}^{\pm }\sqrt{\left( 1-Z^{2K}\right) \left(
1-Z^{2(M-K)}\right) }
\end{eqnarray}
with 
\begin{equation}
Z=\text{ }_{j}\left\langle \alpha \right| \left. -\alpha \right\rangle
_{j}=\exp (-2|\alpha |^{2}),\text{ }\forall j.  \label{Z}
\end{equation}
The concurrence $C_{\{K\},\{M-K\}}^{\pm }$ characterizing the bipartite
entanglement measure between the two subsystems $\{K\}$ and $\{M-K\}$ in the
states $\left| \Psi _{M}^{\pm }\right\rangle $ is then given by 
\begin{eqnarray}
C_{\{K\},\{M-K\}}^{\pm } &=&C_{\{M-K\},\{K\}}^{\pm }=2|a_{00}^{\pm
}a_{11}^{\pm }-a_{01}^{\pm }a_{10}^{\pm }|  \nonumber \\
&=&\frac{\sqrt{\left( 1-\exp (-4K|\alpha |^{2})\right) \left( 1-\exp
[-4(M-K)|\alpha |^{2}]\right) }}{1\pm \exp (-2M|\alpha |^{2})}.
\label{Conct}
\end{eqnarray}
In the limit $|\alpha |\rightarrow \infty $ both $C_{\{K\},\{M-K\}}^{+}$ and 
$C_{\{K\},\{M-K\}}^{-}$ tend to unity, but in the limit $|\alpha
|\rightarrow 0,$ $C_{\{K\},\{M-K\}}^{+}\rightarrow 0$ while $%
C_{\{K\},\{M-K\}}^{-}\rightarrow 2\sqrt{K(M-K)}/M.$ Only the concurrence $%
C_{\{K\},\{M-K\}}^{-}$ can constantly be equal to 1 (maximally entangled)
when the two subsystems become identical, {\it i.e.} when $K=M-K.$ The
dependence of $C_{\{K\},\{M-K\}}^{\pm }$ on $|\alpha |^{2}$ is illustrated
in Fig. 4. The $3$-tangle \cite{coff} and $2N$-tangle \cite{wong} of the
states $\left| \Psi _{M}^{\pm }\right\rangle $ can also be calculated (see, 
{\it e.g.}, Ref. \cite{wangsanders}). It is strictly the entanglement
between modes that make the states $\left| \Psi _{M}^{\pm }\right\rangle $
capable of serving as a quantum channel in quantum teleportation of a
multimode cat-state which is the subject of the next section.

\noindent {\bf 3. Network teleportation}

\noindent Given a network consisting of an arbitrary $N$ parties the task is
to teleport from any one to any another party an unknown multimode cat-state
of the form (\ref{E1}).

\noindent {\it 3.1. Single-mode cat-state}

\noindent The simplest cat-state is with $L=1$ in which case the state (\ref
{E1}) reduces to

\begin{equation}
\left| \Psi \right\rangle _{a}=A_{a}\left( x\left| \alpha \right\rangle
_{a}+y\left| -\alpha \right\rangle _{a}\right)  \label{psia}
\end{equation}
with 
\begin{equation}
A_{a}=\left( |x|^{2}+|y|^{2}+2\exp (-2|\alpha |^{2})\Re (x^{*}y)\right)
^{-1/2}.
\end{equation}
We shall use as a quantum channel the state $\left| \Psi
_{N}^{-}\right\rangle $ which is to be equally shared by all the parties:
each party holds a mode (In the case of $L=1$ the network is naturally
symmetric but for $L>1$ symmetric and asymmetric networks are to be
distinguished as will be seen later). Without loss of generality, we suppose
that party 1 possesses the state $\left| \Psi \right\rangle _{a}$ and the
task is to teleport it to party $N.$ The entire system of the single-mode
cat-state to be teleported, state (\ref{psia}), and the $N$-mode entangled
cat-state of the quantum channel, state $\left| \Psi _{N}^{-}\right\rangle $%
, is

\begin{eqnarray}
\left| \Phi \right\rangle _{a12...N} &=&\left| \Psi \right\rangle _{a}\left|
\Psi _{N}^{-}\right\rangle  \nonumber \\
&=&A_{a}A_{N}^{-}\left( x\left| \alpha \right\rangle _{a}\left| \alpha
\right\rangle _{1}\left| \alpha \right\rangle _{2}...\left| \alpha
\right\rangle _{N}\right.  \nonumber \\
&&-x\left| \alpha \right\rangle _{a}\left| -\alpha \right\rangle _{1}\left|
-\alpha \right\rangle _{2}...\left| -\alpha \right\rangle _{N}  \nonumber \\
&&+y\left| -\alpha \right\rangle _{a}\left| \alpha \right\rangle _{1}\left|
\alpha \right\rangle _{2}...\left| \alpha \right\rangle _{N}  \nonumber \\
&&-\left. y\left| -\alpha \right\rangle _{0}\left| -\alpha \right\rangle
_{1}\left| -\alpha \right\rangle _{2}...\left| -\alpha \right\rangle
_{N}\right) .  \label{prodN}
\end{eqnarray}
At party 1 station, the operator $\widehat{{\cal B}}_{a,1}$ is applied to
mode $a$ and mode $1$ transforming state (\ref{prodN}) into

\begin{eqnarray}
\left| \Theta \right\rangle _{a12...N} &=&A_{a}A_{N}^{-}\left( x\left|
\alpha \sqrt{2}\right\rangle _{a}\left| 0\right\rangle _{1}\left| \alpha
\right\rangle _{2}...\left| \alpha \right\rangle _{N}\right.  \nonumber \\
&&-x\left| 0\right\rangle _{a}\left| \alpha \sqrt{2}\right\rangle _{1}\left|
-\alpha \right\rangle _{2}...\left| -\alpha \right\rangle _{N}  \nonumber \\
&&+y\left| 0\right\rangle _{a}\left| -\alpha \sqrt{2}\right\rangle
_{1}\left| \alpha \right\rangle _{2}...\left| \alpha \right\rangle _{N} 
\nonumber \\
&&-\left. y\left| -\alpha \sqrt{2}\right\rangle _{a}\left| 0\right\rangle
_{1}\left| -\alpha \right\rangle _{2}...\left| -\alpha \right\rangle
_{N}\right) .  \label{TetaN}
\end{eqnarray}
To fulfill the task mentioned above party 1 needs counting the photon
numbers of mode $a$ and mode $1,$ while parties 2, 3, $\ldots ,$ $N-1$
should respectively carry out the local number measurement of their modes.
Let the measurement outcomes at party 1 station be $n_{a}$ and $n_{1},$
whereas $n_{2},$ $n_{3},$ ..., $n_{N-1}$ photons are detected at the
stations of party 2, party 3, ..., party $N-1,$ respectively. The structure
of state $\left| \Theta \right\rangle _{a12...N},$ Eq. (\ref{TetaN}),
excludes the possibility of both $n_{a}\neq 0$ and $n_{1}\neq 0.$ The
outcome $n_{a}=n_{1}=0$ may occur but does not help the task of
teleportation (moreover, its probability depends on $x,y$ and is thus
unknown). We are therefore left only with two possibilities: (i) $n_{a}=0,$ $%
n_{1}>0$ and (ii) $n_{a}>0,$ $n_{1}=0.$ If Case (i) happens then the state
at party $N$ station collapses into

\begin{equation}
\left| \Psi ^{^{(i)}}\right\rangle _{N}=A_{a}\left[
(-1)^{n_{2}+n_{3}+...+n_{N-1}}x\left| -\alpha \right\rangle
_{N}-(-1)^{n_{1}}y\left| \alpha \right\rangle _{N}\right] .  \label{pspN}
\end{equation}
Clearly, if all the parties who performed the number measurement send their
outcomes to party $N$ via a public (classical) channel and the outcomes are
such that $n_{1}+n_{2}+...+n_{N-1}$ is odd, then after obtaining the
classical information party $N$ will apply the operator $\widehat{P}_{N}(\pi
)$ to Eq. (\ref{pspN}) to get the state $\left| \Psi \right\rangle _{N}=%
\widehat{P}_{N}(\pi )\left| \Psi ^{(i)}\right\rangle _{N}$ which is nothing
else but the desired state $\left| \Psi \right\rangle _{a}$ (up to a global
unimportant phase constant sometimes). The probability of success is equal to

\begin{eqnarray}
\Pi _{N}^{(i)} &=&\left| _{1}\left\langle n_{1}\right| _{2}\left\langle
n_{2}\right| ..._{N-1}\left\langle n_{N-1}\right| \left. \Theta
\right\rangle _{a12...N}\right| ^{2}  \nonumber \\
&=&(A_{N}^{-})^{2}\sum_{n_{1}=1,n_{j>1}=0,\sum n_{k}=odd}^{\infty }\left| 
{\cal N}_{n_{1}}(\alpha \sqrt{2}){\cal N}_{n_{2}}(\alpha )...{\cal N}%
_{n_{N-1}}(\alpha )\right| ^{2}.
\end{eqnarray}
Alternatively, if Case (ii) occurs then the state at party $N$ station
collapses into

\begin{equation}
\left| \Psi ^{(ii)}\right\rangle _{N}=A_{a}\left[ x\left| \alpha
\right\rangle _{N}-(-1)^{n_{a}+n_{2}+n_{3}+...+n_{N-1}}y\left| -\alpha
\right\rangle _{N}\right] .  \label{ps2pN}
\end{equation}
Transparently, if $n_{a}+n_{2}+n_{3}+...+n_{N-1}$ is odd, then, by doing
nothing, party $N$ gets an exact replica of the desired state $\left| \Psi
\right\rangle _{a}.$ In this situation the success probability $\Pi
_{N}^{(ii)}=\left| _{a}\left\langle n_{a}\right| _{2}\left\langle
n_{2}\right| ..._{N-1}\left\langle n_{N-1}\right| \left. \Theta
\right\rangle _{a12...N}\right| ^{2}$ is precisely equal to that in Case
(i), {\it i.e}., $\Pi _{N}^{(ii)}=\Pi _{N}^{(i)}.$ Hence, the total
probability of successful teleportation can be calculated explicitly and, as
a result, 
\begin{equation}
\Pi _{N}^{(1-)}=2\Pi _{N}^{(i)}=\frac{1}{2}\left\{ 1-\text{csch}(N|\alpha
|^{2})\sinh \left[ (N-2)|\alpha |^{2}\right] \right\}  \label{PN}
\end{equation}
where the superscript $``(1-)"$ indicates that the teleported cat-state is
single-mode, and the state $\left| \Psi _{N}^{-}\right\rangle $ is used as a
quantum channel. For $N=2$ the formula (\ref{PN}) yields $\Pi
_{2}^{(1-)}=1/2,$ independent of $\alpha ,$ recovering the result reported
in Ref. \cite{enk}. For any $N>2,$ the probability of perfect teleportation
tends to $1/N$ in the limit $|\alpha |\rightarrow 0$ and to $1/2$ in the
limit $|\alpha |\rightarrow \infty $ (see Fig. 5).

\noindent {\it 3.2. Two-mode cat-state}

\noindent We now study the network teleportation problem for a two-mode
cat-state $(L=2)$ of the form

\begin{equation}
\left| \Psi \right\rangle _{ab}=A_{ab}\left( x\left| \alpha \right\rangle
_{a}\left| \alpha \right\rangle _{b}+y\left| -\alpha \right\rangle
_{a}\left| -\alpha \right\rangle _{b}\right)  \label{psiab}
\end{equation}
with 
\begin{equation}
A_{ab}=\left( |x|^{2}+|y|^{2}+2\exp (-4|\alpha |^{2})\Re (x^{*}y)\right)
^{-1/2}.
\end{equation}
For a symmetric network of $N$ parties we use in this circumstance the state 
$\left| \Psi _{2N}^{-}\right\rangle $ as a quantum channel which is equally
shared among all the parties by sending a pair of modes $\{2q-1,2q\}$ to a
party $q$ where $q=1,2,...,N.$ The entire system is described by

\begin{eqnarray}
\left| \Phi \right\rangle _{ab12...2N} &=&\left| \Psi \right\rangle
_{ab}\left| \Psi _{2N}^{-}\right\rangle  \nonumber \\
&=&A_{ab}A_{2N}^{-}\left( x\left| \alpha \right\rangle _{a}\left| \alpha
\right\rangle _{b}\left| \alpha \right\rangle _{1}\left| \alpha
\right\rangle _{2}\left| \alpha \right\rangle _{3}...\left| \alpha
\right\rangle _{2N}\right.  \nonumber \\
&&-x\left| \alpha \right\rangle _{a}\left| \alpha \right\rangle _{b}\left|
-\alpha \right\rangle _{1}\left| -\alpha \right\rangle _{2}\left| -\alpha
\right\rangle _{3}...\left| -\alpha \right\rangle _{2N}  \nonumber \\
&&+y\left| -\alpha \right\rangle _{a}\left| -\alpha \right\rangle _{b}\left|
\alpha \right\rangle _{1}\left| \alpha \right\rangle _{2}\left| \alpha
\right\rangle _{3}...\left| \alpha \right\rangle _{2N}  \nonumber \\
&&+\left. y\left| -\alpha \right\rangle _{a}\left| -\alpha \right\rangle
_{b}\left| -\alpha \right\rangle _{1}\left| -\alpha \right\rangle _{2}\left|
-\alpha \right\rangle _{3}...\left| -\alpha \right\rangle _{2N}\right) .
\label{p2Nt1}
\end{eqnarray}
Obviously, the symmetry allows teleportation from any party $k$ to any other
party $l$ $(l\neq k).$ Without loss of generality, we suppose that party 1
possesses the state $\left| \Psi \right\rangle _{ab}$ and wishes to teleport
it to party $N.$ Then, at party 1 location mode 1 is mixed with mode $b$ and
mode 2 with mode $a$ (or mode 1 with mode $a$ and mode 2 with mode $b,$ the
result will be the same). That is, party 1 applies the operator $\widehat{%
{\cal B}}_{a,2}\widehat{{\cal B}}_{b,1}$ to his/her modes $a,$ $b,$ $1$ and $%
2$ transforming state (\ref{p2Nt1}) into

\begin{eqnarray}
\left| \Theta \right\rangle _{ab12...2N} &=&A_{ab}A_{2N}^{-}\left( x\left|
\alpha \sqrt{2}\right\rangle _{a}\left| \alpha \sqrt{2}\right\rangle
_{b}\left| 0\right\rangle _{1}\left| 0\right\rangle _{2}\left| \alpha
\right\rangle _{3}...\left| \alpha \right\rangle _{2N}\right.  \nonumber \\
&&-x\left| 0\right\rangle _{a}\left| 0\right\rangle _{b}\left| \alpha \sqrt{2%
}\right\rangle _{1}\left| \alpha \sqrt{2}\right\rangle _{2}\left| -\alpha
\right\rangle _{3}...\left| -\alpha \right\rangle _{2N}  \nonumber \\
&&+y\left| 0\right\rangle _{a}\left| 0\right\rangle _{b}\left| -\alpha \sqrt{%
2}\right\rangle _{1}\left| -\alpha \sqrt{2}\right\rangle _{2}\left| \alpha
\right\rangle _{3}...\left| \alpha \right\rangle _{2N}  \nonumber \\
&&-\left. y\left| -\alpha \sqrt{2}\right\rangle _{a}\left| -\alpha \sqrt{2}%
\right\rangle _{b}\left| 0\right\rangle _{1}\left| 0\right\rangle _{2}\left|
-\alpha \right\rangle _{3}...\left| -\alpha \right\rangle _{2N}\right) .
\label{thetaab}
\end{eqnarray}
After that party 1 counts the photon numbers of modes $a,$ $b,$ $1$ and $2,$
while parties $q$ $(q=2,3,...,N-1)$ carry out the local number measurement
of modes $2q-1$ and $2q$ at their location. Let the outcome be $n_{a},$ $%
n_{b},$ $n_{2q-1}$ and $n_{2q}$ $(q=1,2,...,N-1).$ Similar arguments as in
the previous subsection lead to only two situations: (i) $n_{a}=n_{b}=0,$ $%
n_{1}+n_{2}>0$ and (ii) $n_{a}+n_{b}>0,$ $n_{1}=n_{2}=0.$ If Case (i)
happens then the state at party $N$ station collapses into

\begin{equation}
\left| \Psi ^{^{(i)}}\right\rangle _{N}=A_{ab}\left[
(-1)^{n_{3}+...+n_{2(N-1)}}x\left| -\alpha \right\rangle _{2N-1}\left|
-\alpha \right\rangle _{2N}-(-1)^{n_{1}+n_{2}}y\left| \alpha \right\rangle
_{2N-1}\left| \alpha \right\rangle _{2N}\right] .  \label{psp2}
\end{equation}
Clearly, if all the parties 1, 2, ..., $N-1$ send their outcomes to party $N$
via a public (classical) channel and the outcomes are such that $%
n_{1}+n_{2}+...+n_{2(N-1)}$ is odd, then after obtaining the classical
information party $N$ will apply the operator $\widehat{P}_{2N-1}(\pi )%
\widehat{P}_{2N}(\pi )$ to Eq. (\ref{psp2}) to get the state $\left| \Psi
\right\rangle _{N}=\widehat{P}_{2N-1}(\pi )\widehat{P}_{2N}(\pi )\left| \Psi
^{(i)}\right\rangle _{N}$ which is nothing else but the desired state $%
\left| \Psi \right\rangle _{ab}$ (up to a global unimportant phase constant
sometimes). The probability of success is equal to

\begin{eqnarray}
\Pi _{N}^{(i)} &=&\left| _{1}\left\langle n_{1}\right| _{2}\left\langle
n_{2}\right| ..._{2(N-1)}\left\langle n_{2(N-1)}\right| \left. \Theta
\right\rangle _{ab12...2N}\right| ^{2}  \nonumber \\
&=&(A_{2N}^{-})^{2}\sum_{\{n_{j}=0\}}^{\infty }{}^{\prime }\left| {\cal N}%
_{n_{1}}(\alpha \sqrt{2}){\cal N}_{n_{2}}(\alpha \sqrt{2}){\cal N}%
_{n_{3}}(\alpha )...{\cal N}_{n_{2(N-1)}}(\alpha )\right| ^{2}
\end{eqnarray}
with the prime to exclude the term with $n_{1}=n_{2}=0$ and to sum over odd $%
n_{1}+n_{2}+...+n_{2(N-1)}.$ Alternatively, if Case (ii) occurs then the
state at party $N$ station collapses into

\begin{equation}
\left| \Psi ^{(ii)}\right\rangle _{N}=A_{ab}\left[ x\left| \alpha
\right\rangle _{2N-1}\left| \alpha \right\rangle
_{2N}-(-1)^{n_{a}+n_{b}+n_{3}+...+n_{2(N-1)}}y\left| -\alpha \right\rangle
_{2N-1}\left| -\alpha \right\rangle _{2N}\right] .  \label{ps2p}
\end{equation}
Transparently, if $n_{a}+n_{b}+n_{3}+...+n_{2(N-1)}$ is odd, then, by doing
nothing, party $N$ gets an exact replica of the desired state $\left| \Psi
\right\rangle _{ab}.$ In this situation the success probability $\Pi
_{N}^{(ii)}=\left| _{a}\left\langle n_{a}\right| _{b}\left\langle
n_{b}\right| _{3}\left\langle n_{3}\right| ..._{2(N-1)}\left\langle
n_{2(N-1)}\right| \left. \Theta \right\rangle _{ab12...2N}\right| ^{2}$ is
precisely equal to that in Case (i), {\it i.e}., $\Pi _{N}^{(ii)}=\Pi
_{N}^{(i)}.$ The total probability of successful teleportation can be
calculated explicitly as 
\begin{equation}
\Pi _{N}^{(2-)}=2\Pi _{N}^{(i)}=\frac{1}{2}\left\{ 1-\text{csch}(2N|\alpha
|^{2})\sinh \left[ 2(N-2)|\alpha |^{2}\right] \right\}  \label{PN2}
\end{equation}
where the superscript ``$(2-)$'' indicates that the teleported cat-state is
two-mode, and the state $\left| \Psi _{2N}^{-}\right\rangle $ is used as a
quantum channel. As in the case of $L=1$ the success probability is 50\% for 
$N=2$ independent of $\alpha .$ For $N>2$ it is less than but quickly tends
to 50\% with increasing $|\alpha |.$

If network symmetry is not required a less expensive quantum channel can be
used by a $(2N-1)$-mode entangled state, {\it i.e}. by using the state $%
\left| \Psi _{2N-1}^{-}\right\rangle .$ Since symmetry of the network is
broken, the quantum channel cannot be shared by the parties before an
announcement is made about who holds the state $\left| \Psi \right\rangle
_{ab}.$ If party $k$ holds it, one of the $2N-1$ modes of the state $\left|
\Psi _{2N-1}^{-}\right\rangle $ should be sent to party $k$ and each of the
remaining parties receives a pair of modes: an unequal distribution. For
definiteness, we assume that party 1 holds the state $\left| \Psi
\right\rangle _{ab}.$ Then we send mode 1 to party 1 and a pair of modes $%
\{2(q-1),2q-1\}$ to a party $q$ $(q=2,$ $3,$ $\ldots ,$ $N).$ Party 1
applies the operator $\widehat{{\cal B}}_{b,1}$ to modes $b$ and $1$ (or $%
\widehat{{\cal B}}_{a,1}$ to modes $a$ and $1,$ the result will be the
same). All the parties except party $N$ who is supposed to get the
teleported state should measure the photon number of their modes and send
the outcome to party $N.$ Similarly to the above described technique,
depending on the measurement outcome, party $N$ is able to obtain the
desired state $\left| \Psi \right\rangle _{ab}$ by doing nothing or by
applying appropriate operations on his/her state. The total probability of
successful teleportation now is 
\begin{equation}
\widetilde{\Pi }_{N}^{(2-)}=\frac{1}{2}\left\{ 1-\text{csch}[(2N-1)|\alpha
|^{2}]\sinh \left[ (2N-3)|\alpha |^{2}\right] \right\}  \label{Pnga}
\end{equation}
which is always $\alpha $-dependent, even for $N=2.$ Comparing Eq. (\ref
{Pnga}) with Eq. (\ref{PN}) we establish the following relationship

\begin{equation}
\widetilde{\Pi }_{N}^{(2-)}=\Pi _{2N-1}^{(1-)}.  \label{P12}
\end{equation}

\noindent {\it 3.3. Multimode cat-state}

\noindent It is straightforward to generalize the teleportation scenario to
the case of an arbitrary $L$-mode cat-state, i.e. the state (\ref{E1}). The
symmetric network of $N$ parties requires a quantum channel to be served by
a state with $LN$ modes being entangled, i.e. the state $\left| \Psi
_{LN}^{-}\right\rangle $ of which a collection of $L$ modes $\{(q-1)L+1,$ $%
(q-1)L+2,$ $\ldots ,$ $qL\}$ is sent to a party $q$ $(1\leq q\leq N).$ Note
again that this equal sharing of the quantum channel can be done at any time
before an actual teleportation takes place. For teleporting from a party,
say, party 1, to another party, say, party $N,$ party 1 mixes his/her modes
pairwise: one from the group $\{a_{1},$ $a_{2},$ $\ldots ,$ $a_{L}\}$ and
one from the group $\{1,$ $2,$ $\ldots ,$ $L\}.$ Then the parties 1, 2, $%
\ldots ,$ $N-1$ carry out appropriate number measurement and communicate the
outcome to party $N$ who will get the teleported state by the right action
as mentioned above. The explicit expression of success probability in this
general case can also be derived. It is

\begin{equation}
\Pi _{N}^{(L-)}=\frac{1}{2}\left\{ 1-\text{csch}(LN|\alpha |^{2})\sinh
\left[ L(N-2)|\alpha |^{2}\right] \right\}  \label{PL}
\end{equation}
where the superscript ``$(L-)$'' indicates that the teleported cat-state is $%
L$-mode, and the state $\left| \Psi _{LN}^{-}\right\rangle $ is used as a
quantum channel. Generally $\Pi _{N}^{(L-)}$ depends on $N,$ $L$ and $\alpha
,$ but for $N=2$ it is exactly $50\%$ independent of both $L$ and $\alpha .$

In the case of asymmetric network we can use a less expensive quantum
channel of only $(L(N-1)+1)$ modes, {\it i.e.} the state $\left| \Psi
_{L(N-1)+1}^{-}\right\rangle .$ Now only after knowing who holds the state
to be teleported the parties start to share the quantum channel. Let the
state $\left| \Psi \right\rangle _{a_{1}a_{2}...a_{L}}$ be with party 1 and
party $N$ receive it. Then, mode 1 should be sent to party 1 and a
collection of $L$ modes $\{(q-2)L+2,$ $(q-2)L+3,$ $\ldots ,$ $qL+1\}$ to a
party $q=2,3,...,N:$ again an unequal sharing. After that party 1 applies
the operator $\widehat{{\cal B}}_{a_{j},1}$ to modes $a_{j}$ and $1$ where $%
a_{j}$ is any one among the modes $\{a_{1},$ $a_{2},$ $\ldots ,$ $a_{L}\}.$
All the parties except party $N$ measure the photon number of their modes
and classically communicate their outcomes with party $N.$ Depending on the
communication content, party $N$ automatically get the desired cat-state $%
\left| \Psi \right\rangle _{a_{1}a_{2}...a_{L}}$ or is able by applying
proper operations to convert the state in his/her location into $\left| \Psi
\right\rangle _{a_{1}...a_{L}}.$ The total probability of success within
such an asymmetric network is related to $\Pi _{M}^{(1-)}$ as

\begin{equation}
\widetilde{\Pi }_{N}^{(L-)}=\Pi _{(N-1)L+1}^{(1-)}  \label{PLN}
\end{equation}
which always depends on $\alpha ,$ even for $N=2.$

\noindent {\it 3.4. Discussion}

\noindent As seen from the analysis, for a symmetric network the probability
of successful teleportation reaches 50\% when $N=2$ independent of both $L$
and $\alpha ,$ {\it i.e.} $\Pi _{2}^{(L-)}=50\%$ no matter how many modes
the state to be teleported is composed of and how intense the applied field
is. For an asymmetric network an entangled state of lesser modes can be used
as a quantum channel but the probability of success depends on $\alpha $ for
all $N$ including $N=2.$ Generally for both symmetric and asymmetric
networks the success probability quickly saturates to 50\%. In fact, as Fig.
5 shows, the probability already reaches 50\% starting from $|\alpha |^{2}=3$
for which $\overline{n}^{\pm }$ well approaches $|\alpha |^{2}$ too (see Eq.
(\ref{ntbct}) and Fig. 3). So, in a sense, for fields with an averaged
photon number per mode equal to or greater than 3 the success of
teleportation can be considered as being taken place with a probability
50\%, independent of $\alpha ,$ $L$ and $N$ as well. However, the symmetric
network has an advantage over the asymmetric one in the sense that the
quantum channel can be shared equally among the participants at any time
before the teleportation process and the network is ready for teleporting
between any two locations at a later time.

In the proposed teleportation protocols state-parities rather than photon
numbers themselves are needed. Thus the kind of required measurement remains
the same as for preparation of the states $\left| \Psi _{M}^{\pm
}\right\rangle ,$ {\it i.e}. they are parity measurements. These can easily
be performed, say, by coupling the Fock state appearing after a number
measurement to a two-level atom via a dispersive interaction (see, {\it e.g.}%
, Ref. \cite{scully}) governed by the interaction Hamiltonian $%
H_{int}=fa_{j}^{+}a_{j}\sigma _{x}$ with $f$ the coupling strength and $%
\sigma _{x}=\left| g\right\rangle \left\langle e\right| +\left|
e\right\rangle \left\langle g\right| $ $(\left| g\right\rangle $ $(\left|
e\right\rangle )$ the atomic ground (excited) state). If the atom is
initially prepared either in the excited or in the ground state and is to be
detected after an interaction time $\tau $ such that $\tau f=\pi /2,$ then a
state-flipping ({\it e.g.}, $\left| g\right\rangle \rightarrow \left|
e\right\rangle $ or $\left| e\right\rangle \rightarrow \left| g\right\rangle
)$ means an odd parity, whereas no-flipping indicates an even parity, no
matter how many photons in the Fock state are.

In Refs. \cite{zheng,jeong} the nonorthogonal cat-states have been
reconstructed in orthogonal bases and Bell-like state measurements are to be
performed for a teleportation task. However, unlike the original Bell states
discriminating Bell-like states is not trivial, and an experimental setup
was suggested in \cite{jeong}. The setup is probabilistic and does not allow
a perfect discrimination for the Bell-like states. Here we are dealing with
the cat-states as they are, and the necessary teleportation procedures prove
quite simple and natural.

For the same teleportation task we could alternatively use the state $\left|
\Psi _{M}^{+}\right\rangle $ as a quantum channel. Going along similar line
we have derived formulae for the probability of successful teleportation as

\begin{equation}
\Pi _{N}^{(L+)}=\frac{1}{2}\left\{ 1-\text{sech}(LN|\alpha |^{2})\cosh
[L(N-2)|\alpha |^{2}]\right\}
\end{equation}
for the symmetric network, and

\begin{equation}
\widetilde{\Pi }_{N}^{(L+)}=\Pi _{(N-1)L+1}^{(1+)}
\end{equation}
for the asymmetric network with $\Pi _{M}^{(1+)}$ for any $M\geq 2$ given by

\begin{equation}
\Pi _{M}^{(1+)}=\frac{1}{2}\left\{ 1-\text{sech}(M|\alpha |^{2})\cosh
[(M-2)|\alpha |^{2}]\right\} .
\end{equation}
Although both $\Pi _{N}^{(L+)}$ and $\widetilde{\Pi }_{N}^{(L+)}$ saturate
to 1/2 for large $|\alpha |,$ they tend to vanish in the limit $|\alpha
|\rightarrow 0$ (see the dashed curves in Fig. 4). This gives an advantage
of the state $\left| \Psi _{M}^{-}\right\rangle $ over the state $\left|
\Psi _{M}^{+}\right\rangle $ in teleportation utilizing low intensity fields.

It should also be kept in mind that in the proposed protocols the
teleportation cannot be reliable ({\it i.e}. with 100\% probability of
success). There may happen that while using $\left| \Psi
_{M}^{-}\right\rangle $ $(\left| \Psi _{M}^{+}\right\rangle )$ as a quantum
channel the total state-parity of the measured modes turns out to be even
(odd) in which case the state at party $N$ station differs from the original
state $\left| \Psi \right\rangle _{a_{1}a_{2}...a_{L}}$ by a relative phase
factor. Since no unitary transformation that casts $x\left| \alpha
\right\rangle _{a_{1}}\left| \alpha \right\rangle _{a_{2}}...\left| \alpha
\right\rangle _{a_{L}}-y\left| -\alpha \right\rangle _{a_{1}}\left| -\alpha
\right\rangle _{a_{2}}...\left| -\alpha \right\rangle _{a_{L}}$ into $%
x\left| \alpha \right\rangle _{a_{1}}\left| \alpha \right\rangle
_{a_{2}}...\left| \alpha \right\rangle _{a_{L}}+y\left| -\alpha
\right\rangle _{a_{1}}\left| -\alpha \right\rangle _{a_{2}}...\left| -\alpha
\right\rangle _{a_{L}}$ for any $\alpha $ has been available, the
teleportation fails.

\noindent {\bf 4. Conclusion}

\noindent In conclusion, we have proposed two schemes to generate general
multimode entangled cat-states and have then used them to teleport within a
network of $N$ parties an unknown state which is also a kind of multimode
entangled cat-state. If the state to be teleported is $L$-mode the quantum
channel state needs to be $NL$-mode for a symmetric network and $(L(N-1)+1)$%
-mode for an asymmetric network. Both the proposed generation schemes and
teleportation protocols require only linear optical devices (beam-splitters)
and simple parity measurements. The analytical results show that for medium
and high intensity fields $(|\alpha |^{2}\geq 3)$ perfect teleportation can
be achieved with a probability 50\% for any $L$ and $N.$ The teleportation
protocols meet all the requirements imposed by no-cloning theorem and
special relativity. In fact, no more than one party can get an exact replica
of the teleported state because except party $N$ all the other parties
counted the photon number of their modes and thus destroyed the original
state at their locations. Also, no party stands outside the game. Without
participation of any of the parties the teleportation can never be completed
since the measurement outcomes of all the $N-1$ parties should be collected
at party $N$ before that latter party is able to infer the teleported state.
Since the collection of measurement outcome takes a finite time via public
communication superluminal signaling does not occur.

\noindent {\bf Acknowledgments}

Communication with X. Wang is gratefully acknowledged. The work of B.A.N. is
supported by KIAS Research Fund No. 02-0149-001 and that of J.K. by Korea
Research Foundation Grant No. KRF-2002-070-C00029.

\newpage

\begin{center}
{\bf Figure captions}
\end{center}

\begin{enumerate}
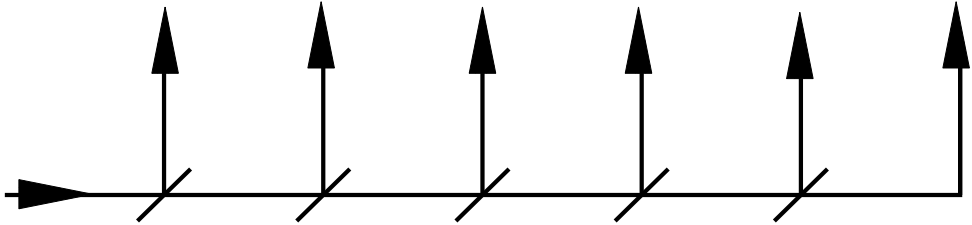
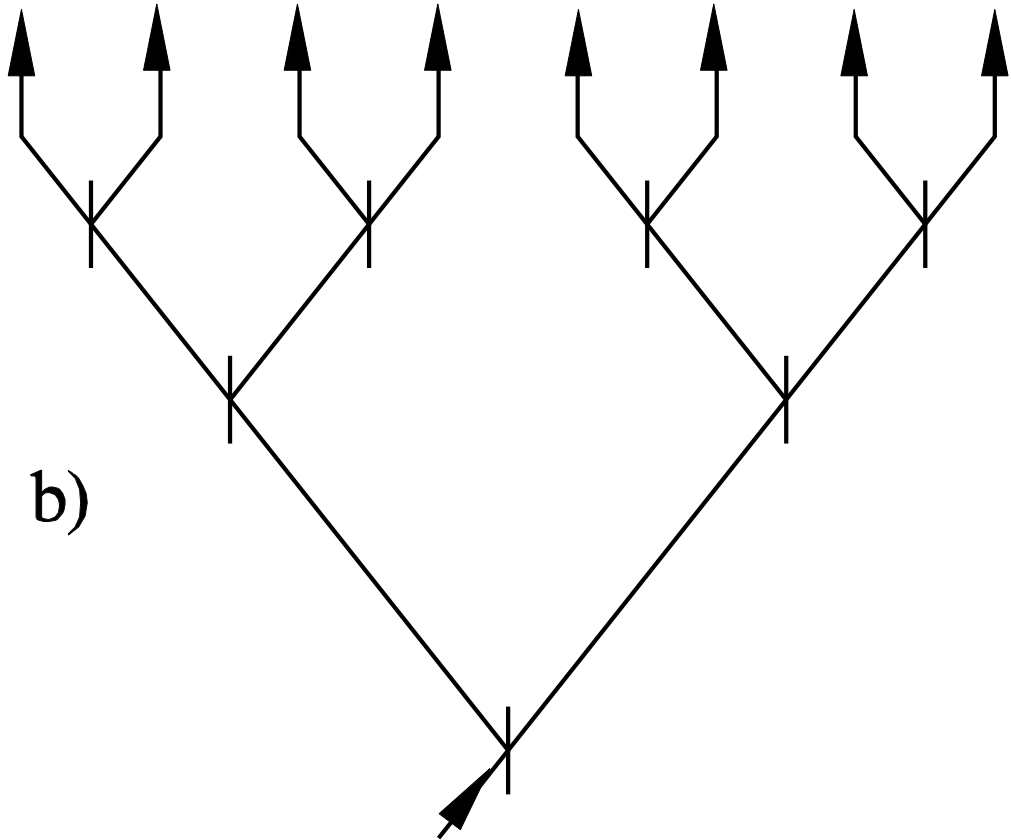
\item[Fig. 1:]  Arrangement of beam-splitters in different schemes. a)
Ladder-scheme for generating state $\left| \Psi _{M}^{\pm }\right\rangle $
from the input in the form of a single-mode cat-state $A_{M}^{\pm }(\left|
\alpha \sqrt{M}\right\rangle \pm \left| -\alpha \sqrt{M}\right\rangle )$
with $M=6.$ Each 45$^{\circ }$-inclined bar represents a beam-splitter $%
\widehat{B}_{k,l}(\theta )$ defined by Eq. (\ref{B}). b) Tree-scheme for
generating state $\left| \Psi _{2^{Q}}^{\pm }\right\rangle $ from the input
in the form of a single-mode cat-state $A_{2^{Q}}^{\pm }(\left| \alpha \sqrt{%
2^{Q}}\right\rangle \pm \left| -\alpha \sqrt{2^{Q}}\right\rangle )$ with $%
Q=3.$ Each vertical bar represents a beam-splitter $\widehat{{\cal B}}_{k,l}$
defined by Eq. (\ref{BN}).

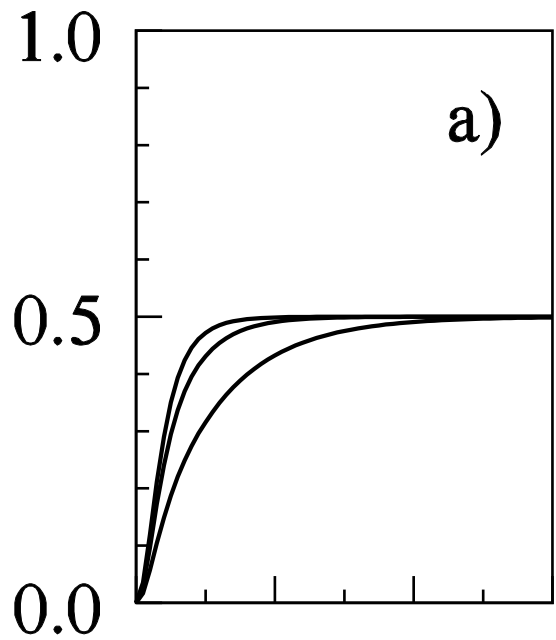
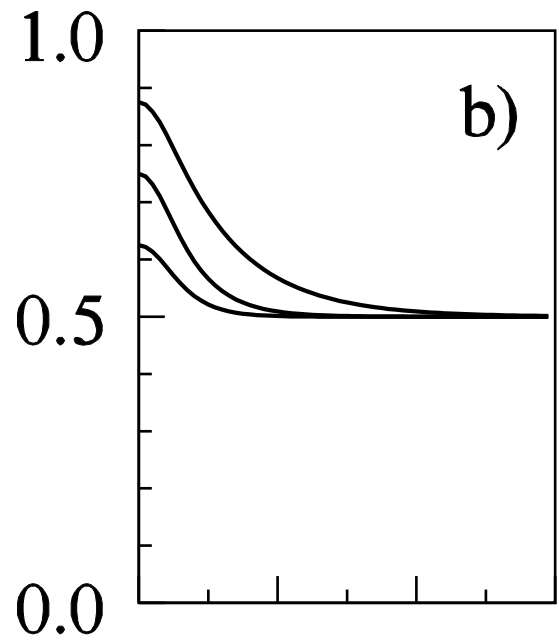
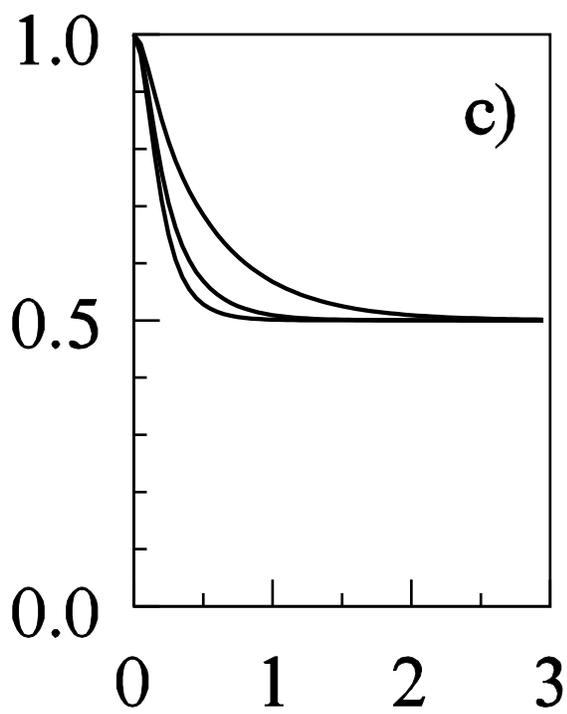
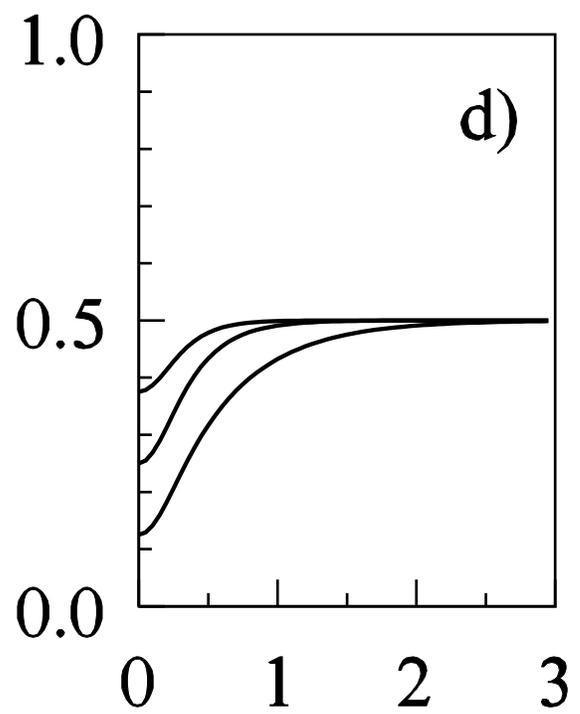
\item[Fig. 2:]  The probabilities a) $P_{Q,M}^{+-}$ (downwadrs), b) $%
P_{Q,M}^{--}$ (upwards), c) $P_{Q,M}^{++}$ (upwards) and d) $P_{Q,M}^{-+}$
(downwards) as a function of $|\alpha |^{2}$ for $Q=3$ and $M=5,6,7.$

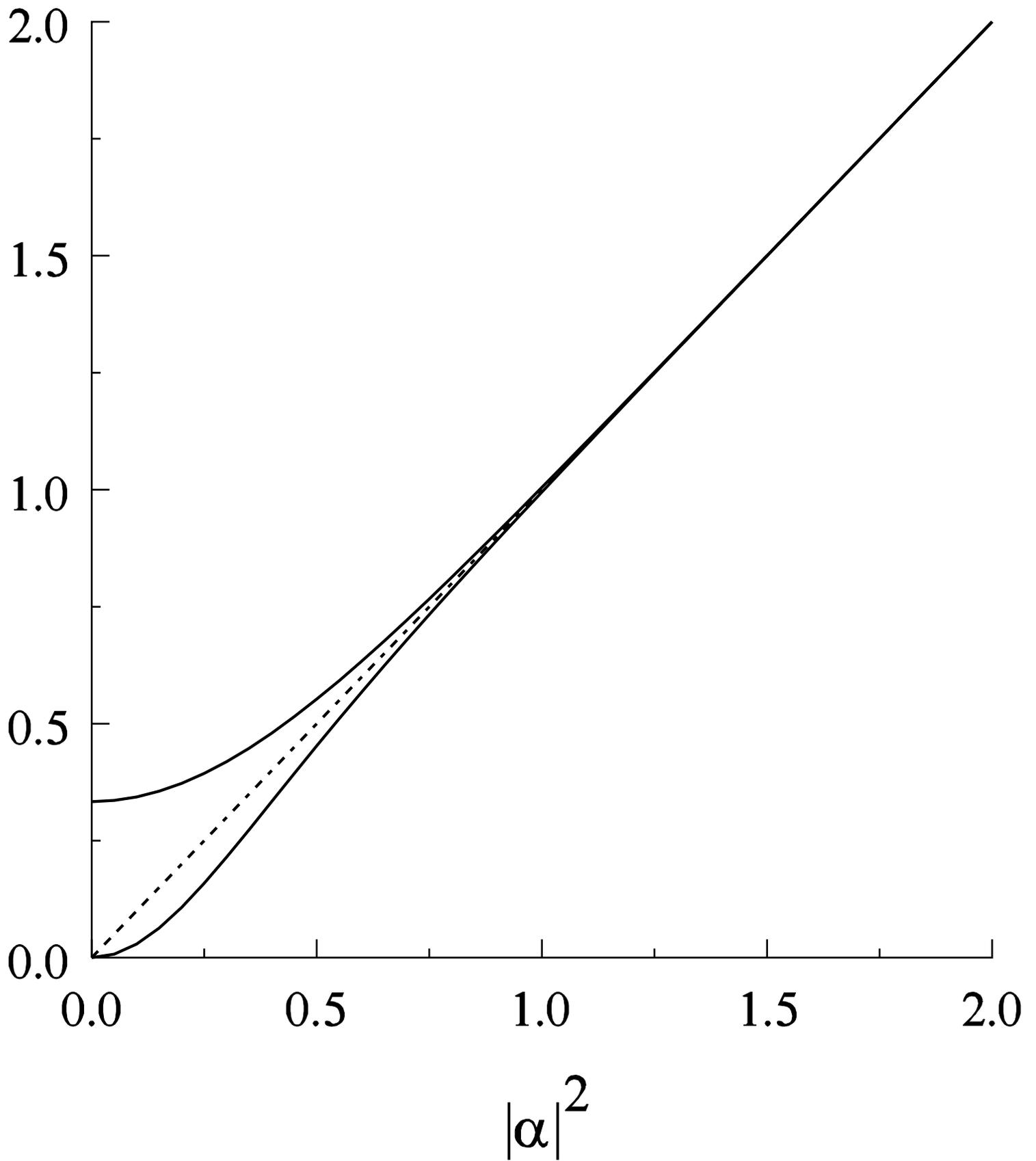
\item[Fig. 3:]  The averaged photon number per mode $\overline{n}^{-}$
(upper solid curve) and $\overline{n}^{+}$ (lower solid curve) in dependence
on $|\alpha |^{2}$ for $M=3.$ The dashed line represents $|\alpha |^{2}$
which is drawn for comparison.

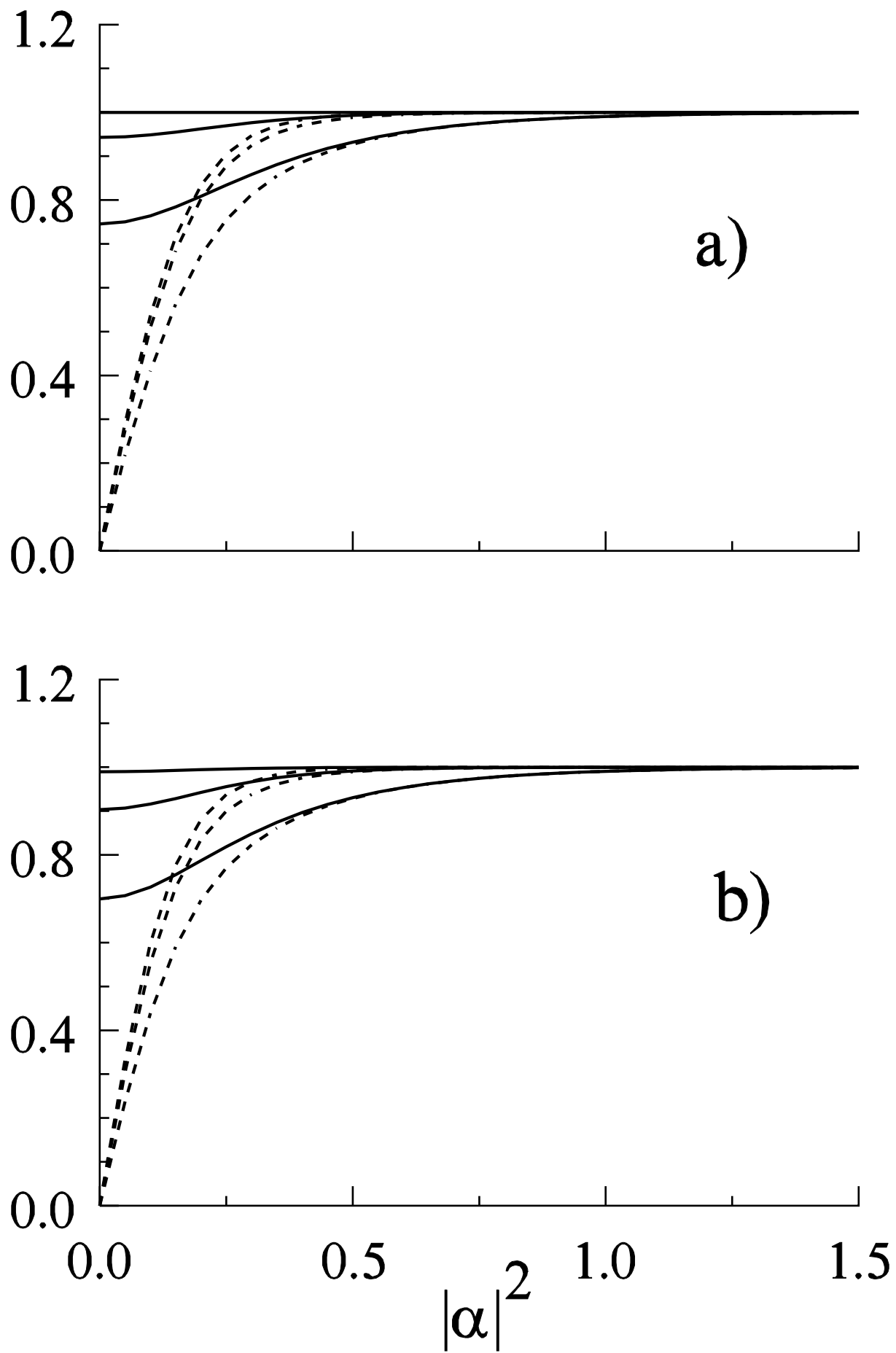
\item[Fig. 4:]  The concurrences $C_{\{K\},\{M-K\}}^{-}$ (solid curves,
upwrads) and $C_{\{K\},\{M-K\}}^{+}$ (dashed curves, upwrads) versus $%
|\alpha |^{2}$ for a) $M=6,$ $K=1(5),$ $2(4)$ and $3,$ and b) $M=7,$ $%
K=1(6), $ $2(5)$ and $3(4).$

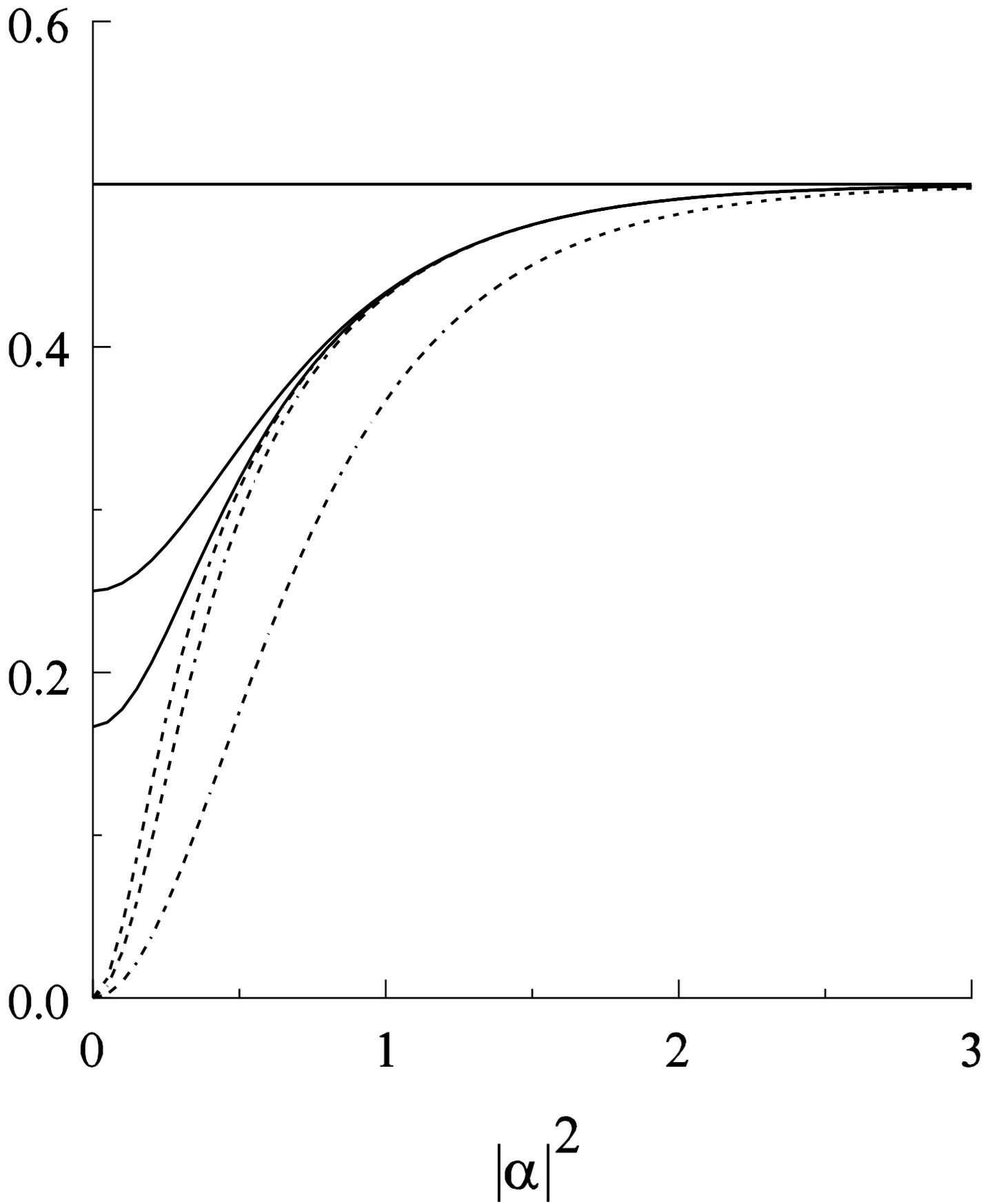
\item[Fig. 5:]  The probability of successful teleportation $\Pi _{N}^{(1-)}$
(solid curves, downwards) and $\Pi _{N}^{(1+)}$ (dashed curves, upwards) as
a function of $|\alpha |^{2}$ for $N=2,$ $4$ and $6.$
\end{enumerate}

\end{document}